\documentclass[a4paper]{article}
\usepackage[margin=25mm]{geometry}
\usepackage{amssymb}

\usepackage{amsmath}
\usepackage{url}
\usepackage{subfigure}
\usepackage{graphicx, xcolor}
\usepackage{amsfonts}
\usepackage[normalem]{ulem}
\usepackage{cite}

\usepackage{fancyhdr}

\usepackage{cite}

\usepackage{color}
\definecolor{darkgreen}{rgb}{0,0.5,0}

\newcommand{\sai}[2][]{%
   {\color{red}\sout{#2}}%
   \ifx&#1&%
      {}
    \else%
      \footnote{{\color{red}(sai):} #1}
    \fi%
}

\begin{document}
%
\title{Approximating Network Centrality Measures Using Node Embedding and Machine Learning}

\author{Matheus R. F. Mendon\c{c}a,
        Andr\'{e} M. S. Barreto, and
        Artur Ziviani
\thanks{M. R. F. Mendon\c{c}a, A. M. S. Barreto and A. Ziviani are with the National Laboratory for Scientific Computing (LNCC), Petr\'{o}polis, RJ, Brazil. E-mails: \{mrfm,amsb,ziviani\}@lncc.br}
\thanks{M. R. F. Mendon\c{c}a is currently with DASA, S\~{a}o Paulo, SP, Brazil and A. M. S. Barreto is currently with DeepMind, London, UK.}}

\date{}



\maketitle

\begin{abstract}
Extracting information from real-world large networks is a key challenge nowadays. For instance, computing a node centrality may become unfeasible depending on the intended centrality due to its computational cost. One solution is to develop fast methods capable of approximating network centralities. Here, we propose an approach for efficiently approximating node centralities for large networks using Neural Networks and Graph Embedding techniques. Our proposed model, entitled Network Centrality Approximation using Graph Embedding (NCA-GE), uses the adjacency matrix of a graph and a set of features for each node (here, we use only the degree) as input and computes the approximate desired centrality rank for every node. NCA-GE has a time complexity of $O(|E|)$, $E$ being the set of edges of a graph, making it suitable for large networks. NCA-GE also trains pretty fast, requiring only a set of a thousand small synthetic scale-free graphs (ranging from 100 to 1000 nodes each), and it works well for different node centralities, network sizes, and topologies. Finally, we compare our approach to the state-of-the-art method that approximates centrality ranks using the degree and eigenvector centralities as input, where we show that the NCA-GE outperforms the former in a variety of scenarios.
\end{abstract}


\section{Introduction}

Networks are present in several real-world applications spread among different disciplines, such as biology, mathematics, sociology, and computer science, just to name a few.  Therefore, network analysis is a crucial tool for extracting relevant information. However, this analysis may be hindered when performed over large complex networks, given the high computational cost for some network analysis methods. This problem is especially challenging when we consider that nowadays many networks extracted from real-world applications are usually large in scale.

Among the most used approaches for network analysis, node centrality is an essential concept related to the assessment of the relative importance of nodes given some criteria of importance. There are several different centrality measures, each with their own definition of a central (i.e., important) node. For example, the most basic node centrality is the degree centrality which measures the number of connections a node has as a proxy for its relative importance in the network. In this case, a node is considered central when it has many connections. Other examples of common  centrality measures are \textit{betweenness}, \textit{closeness}, and \textit{eigenvector} centralities, among others. Centrality measures have already been successfully applied to many real-world problems, as for example: (i)~optimization of computer networks~\cite{app_cn_3}; (ii)~transportation networks~\cite{malha_aerea}; (iii)~biological networks~\cite{app_b_2}; (iv)~recommendation systems~\cite{pinsage}; (v)~reinforcement learning~\cite{graph_rl_survey}; and (vi)~social networks~\cite{app_sn_2}, to name a few domains of application. For more applications, refer to \cite{grando_3,survey_social_net} for a more in-depth review of real-world examples of network centrality usage.

The size of many complex networks found in real-world applications is large. Depending on the size of a network, the direct calculation of a given network centrality measure may become unfeasible due to the required high computational cost. Therefore, many solutions have been proposed for approximating or even calculating the actual centrality value for large networks in a reasonable time, given some constraint or specific target. The initial solutions focused on sampling techniques used for static graphs~\cite{sample_between,sample_close,sample_parameters,aprox_close} or parallel algorithms for specific centrality measures~\cite{parallel_cent}. More recently, researchers have focused on the problem of approximating network centrality for evolving graphs~\cite{aprox_evolve_1}. Another recent approach is to use a Neural Network~(NN) to predict node centrality based on lower-order node features~\cite{grando_1,grando_2,grando_3}, which is the approach that inspired our work.

In this paper, we focus on the problem of approximating node centrality measures for large complex networks using node embedding and Machine Learning. Node~(graph) embedding~\cite{survey_1,survey_3} comprises methods that use neural networks in order to build an abstract representation of a node or a graph, encoding higher-order representations in these low dimension embeddings~(check Section~\ref{sec:related} for more details). We propose here the Network Centrality Approximation using Graph Embedding (NCA-GE) model, a model capable of predicting the rank of the desired centrality for all nodes in a graph of any dimension and topology. To achieve this, our model only requires the adjacency matrix and a set of features for each node in order to predict its value for other centrality measures, whereas other node features are inferred by the node embedding. The set of features reported in this work is composed simply of the degree centrality of all nodes, but any other node feature is allowed. We train our model over a set of artificial complex networks. The model can be thus used to predict the network centrality value of a given node in larger networks. 

The main contributions of the NCA-GE model with respect to other similar previous works~\cite{grando_1,grando_2,grando_3} are:

\begin{itemize}
	\item The NCA-GE model constitutes a more general framework that works with any graph embedding method, making it highly customizable since it can harness the advantages of different graph embedding methods in order to adapt the model for different network topologies.
	\item The NCA-GE model, as investigated here, requires less node features since it uses only the node degree as a feature, whereas the current state-of-the-art model~\cite{grando_3} uses the degree and the eigenvector centrality to this end.
	\item The NCA-GE model outperforms the current state-of-the-art model~\cite{grando_3} by providing more accurate centrality rankings for different network topologies.
\end{itemize}

\noindent The source code for the all the presented models is publicly available on GitHub.\footnote{https://github.com/MatheusMRFM/NCA-GE}

The remainder of this paper is organized as follows. In Section~\ref{sec:basic}, we present the basic concepts of node centrality, node embedding, and neural networks, which compose the foundations of this work. We then present some of the related work on node embedding and network centrality approximation in Section~\ref{sec:related}. In Section \ref{sec:method}, we introduce our proposed NCA-GE model, followed by the description of the experiments and obtained results in Section~\ref{sec:result}. Finally, we conclude the paper and discuss some future work in Section~\ref{sec:conclusion}.

\section{Node Centrality Measures}
\label{sec:basic}

In this section, we briefly describe some important network centrality measures that are widely used and that we consider in this work. In this context, a network is typically represented by a graph $G = (V, E)$, where $V$ is the set of vertices (nodes) and $E$ is the set of edges in the graph. The number of vertices (nodes) in the graph is $N=|V|$. The graph's connectivity is represented by an adjacency matrix $A$, where $a_{ij} \in A$ is the weight of the edge that connects nodes $i$ and $j$. For unweighted graphs, $a_{ij} = 1$ if there exists an edge between nodes $i$ and $j$, and $a_{ij} = 0$ otherwise. For the scope of this work, we consider unweighted and undirected graphs.

There are several ways to define the relative importance of a given node in a graph. This is done using centrality measures: A score given to each node that measures its relative importance in a given context; thus leading to a node ranking. 

\subsubsection{Degree Centrality} 

The degree $d_i$ of node $i$ is given by $d_i = \sum_{j \in V} a_{ij}$. The degree centrality allows the identification of highly connected nodes and represents the most simple centrality measure. However, this centrality is insufficient to explain certain node properties, leading to the definition of more meaningful centrality measures.

\subsubsection{Eigenvector Centrality}

In contrast to the degree centrality, in which each node contributes equally to the centrality of its neighbors, in the eigenvector centrality, central nodes contribute more to the centrality of their neighbor nodes. The eigenvector centrality of a node is thus proportional to the sum of the eigenvector centralities of its neighbor nodes. Therefore, central nodes are usually connected to other central nodes or are connected to a relatively large number of low centrality nodes. The most central nodes typically have both. This is useful to identify influential nodes in a network or nodes connected to other influential nodes. The eigenvector centrality $x_i$ of node $i$ is formally defined as:

\begin{equation}
x_i = \frac{1}{\lambda} \sum_{j \in V} a_{ij} x_j,
\end{equation}

\noindent
where $\lambda$ is an eigenvalue of the adjacency matrix $A$ associated with the eigenvector $x$. Therefore, the eigenvector centrality of each node in a graph can be calculated by solving the eigenvector problem

\begin{equation}
Ax = \lambda x.
\end{equation}


The two centralities presented so far are connectivity-based centralities, i.e., they use node connectivity to measure its importance. Connectivity-based centralities, which are not limited to the two previously presented centralities, are computationally less expensive to be calculated. 

We now present some shortest-path based centralities, which use minimum paths between nodes in order to measure their importance. Hence, these centralities are far more expensive computationally, since they require computing the minimum distance between all pairs of nodes. Check Grando et al.~\cite{grando_3} for more details in the computational time complexity for different types of node centralities.

\subsubsection{Closeness Centrality}

The closeness centrality of a node is given by the mean distance from this node to all other nodes in the graph. The formal definition of closeness centrality for node $i$ is the inverse of the mean minimum distance from that node to all other $N-1$ nodes in the network, given by

\begin{equation}
c_i = \frac{N - 1}{\sum_{j \neq i \in V} \delta(i,j)},
\end{equation}

\noindent where $\delta(i,j)$ is the distance between nodes $i$ and $j$. A high closeness centrality means that the node is on average closer to all other nodes in the network, i.e., the node with high closeness centrality can on average reach other nodes or be reached by these nodes in fewer steps.

\subsubsection{Harmonic Centrality}

Harmonic centrality~\cite{harmonic} represents a variation of the closeness centrality. Instead of calculating the inverse of the overall sum of minimum distances from one node $i$ to all other nodes, harmonic centrality computes the sum of the inverse minimum distances, i.e.

\begin{equation}
h_i = \sum_{j \neq i \in V}\frac{1}{\delta(i,j)},
\end{equation}

\noindent where $1/\delta(i,j)=0$ if there is no path between $i$ and~$j$. Note that this allows infinity distances (for unconnected graphs), which is not allowed for closeness centrality.

\subsubsection{Betweenness Centrality}

In some scenarios, it is desirable to identify nodes by which many shortest paths pass through. This is achieved by the betweenness centrality. In other words, the betweenness centrality of node~$i$ measures the relation of the number of shortest paths~(between any two nodes) that passes through a specific node $i$ and the total number of shortest paths connecting all node pairs.
The betweenness centrality of node $i$ is then formally given by:

\begin{equation}
b_i = \sum_{s \neq i \neq t \in V} \frac{\sigma_{st}(i)}{\sigma_{st}},
\end{equation}

\noindent where $\sigma_{st}(i)$ is the number of shortest paths between nodes $s$ and $t$ that passes through $i$ and $\sigma_{st}$ is the total number of shortest paths between $s$ and $t$. Therefore, betweenness centrality measures the extent to which a node lies on paths between other nodes. Nodes with high betweenness may have considerable influence within a network due to their control over information passing between other nodes.

\section{Related Work}
\label{sec:related}

\begin{figure*}[!t]
	\centering
	\subfigure[GCN Node Embedding.]{\includegraphics[scale=1.0]{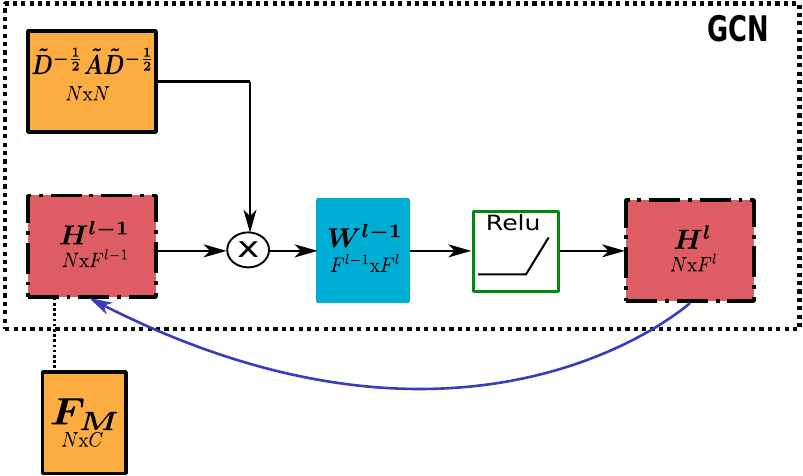}}
	\hspace{1cm}
	\subfigure[Structure2Vec Node Embedding.]{\includegraphics[scale=1.0]{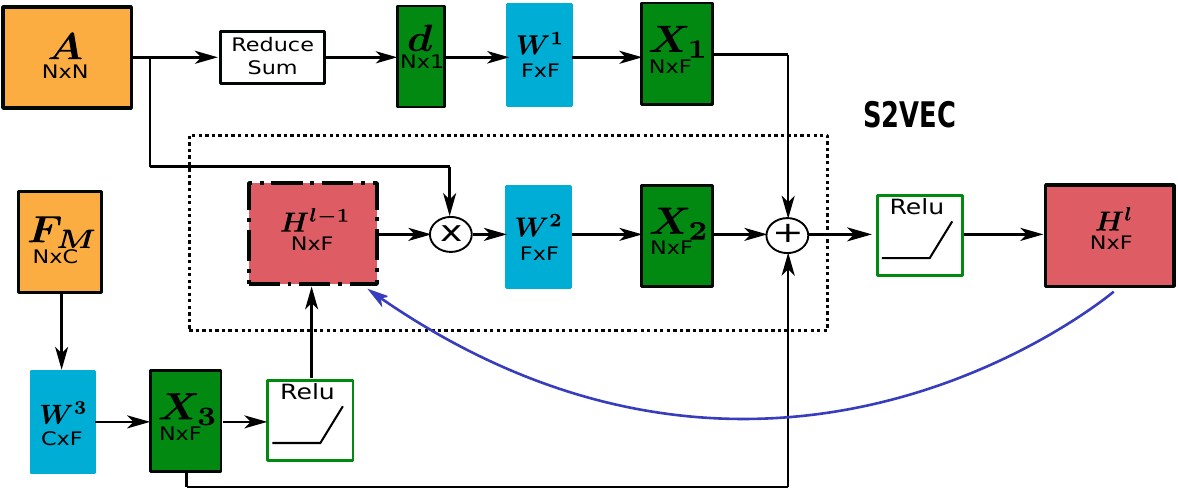}}
	\subfigure[NCA-GE.]{\includegraphics[scale=1.0]{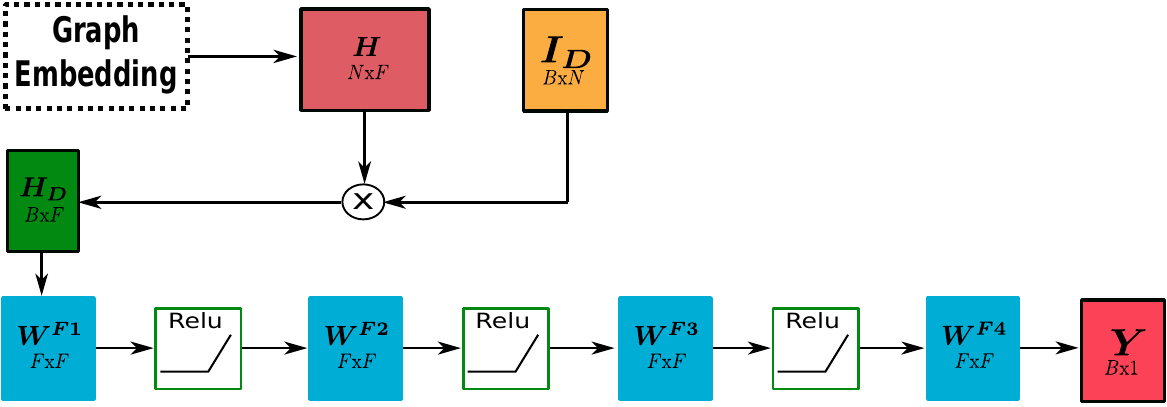}}
	\caption{Architecture used for the graph embedding methods and the proposed model: (a) Graph Convolutional Network (GCN), (b) Structure2Vec (S2VEC), and the Network Centrality Approximation using Graph Embedding~(NCA-GE) model. The inputs are represented in yellow, the node embedding matrix is in light red, hidden  layers are in blue, intermediate matrices are in green, and the output is represented in red. $N$ represents the number of nodes in the graph being processed in the batch, $C$ is the number of features used in the feature matrix, $F$ is the number of features used in the embedding, and $B$ is the batch size. For each batch, we require the normalized adjacency matrix $\Tilde{D}^{-\frac{1}{2}} \Tilde{A} \Tilde{D}^{-\frac{1}{2}} H^{(l)} W^{(l)}$ matrix (for the GCN) or the common adjacency matrix $A$ (for the S2VEC), the feature matrix containing the features of all nodes, and the node id vector containing the id of all nodes in the batch. The output is a single value for each node in the batch, representing the normalized rank of the desired centrality. The proposed model works with any graph embedding method, showcased by the generic ``Graph Embedding" box at the to left corner, which outputs the embedding matrix.}
	\label{fig:architecture}
\end{figure*}

In this section, we briefly review some of the recent advances in node centrality prediction and node embedding, which are the techniques that lay the foundation for this work.

\subsection{Node and Graph Embedding}


Node and graph embedding allows encoding certain node characteristics and topological information into low dimension vectors. These methods usually require two matrices: the adjacency matrix and a feature matrix, containing a set of features for each node. With these two matrices, graph embedding methods use a set of weight matrices in order to learn a special mapping for each node in a higher-order dimension~$F$. Therefore, each node is mapped to an $F$ dimensional vector that reflects each node's features (which depends on the features passed to the model as input) and their connectivity properties, extracted from the adjacency matrix. These weight matrices are learned through an iterative supervised learning process, where in each step (i)~the graph embedding maps each node to their embeddings, which are then (ii)~used to separate each node in specific classes or predict other properties of each node (this depends on the desired task being performed). We can then (iii)~compute the error associated with these predictions (since we are considering a supervised setting) and the gradient of this error function in relation to each weight. These gradients are then (iv)~used to update each weight matrix through the backpropagation algorithm.

There are several different embedding methods, which can be separated into different classes~\cite{survey_1}: (i)~methods based on factorization; (ii)~methods based on random walks; and (iii)~methods based on deep learning. For this work, we focus on methods based on deep learning. The interested reader may refer to one of the many recent surveys on node and graph embedding~\cite{survey_1,survey_3} for a more in-depth analysis of this area.

Wang et al.~\cite{sdne} present the \textit{Structural Deep Network Embedding} (SDNE), a semi-supervised model that builds node embeddings that minimize the First and Second-Order proximity measures. The former refers to the proximity between two nodes based on the weight of the edge connecting them. The latter defines the similarity between two nodes based on their neighborhoods, where two nodes are similar when their neighborhoods are similar. Therefore, SDNE uses two modules in order to build node embeddings that obey these two proximity measures: an unsupervised and a supervised model. The unsupervised model is comprised of two autoencoders that receive a vector $x_i, x_j \in \rm I\!R^N$ for nodes $i$ and $j$, where $N$ is the number of vertices. The vector $x_i$ is actually the row $i$ of the graph's adjacency matrix. These autoencoders learn how to map the neighborhood of each node into a latent state representation within their hidden layers that consider the information of each node's neighborhood. The supervised model, in its turn, uses the latent representation of the autoencoders and minimizes the distance between the latent representation of nodes that are connected in the graph. This allows the model to learn a node embedding that minimizes the first and second-order proximity at the same time.

Graph Convolutional Networks (GCN)~\cite{gcn} represent another node embedding method that uses the data and its underlying graph structure to build powerful latent representations of a node. This architecture receives a matrix $F_M \in \rm I\!R^{NxC}$ of $C$ features for each of the $N$ nodes as an input and outputs a matrix $H \in \rm I\!R^{NxF}$ of $F$ higher-order features for each node, which can be the label of each node, for example ($F$ being the number of possible labels). Each layer $l+1$ of the neural network computes an intermediate matrix $H^{l+1}$ given by:

\begin{equation}
H^{l+1} = \sigma \left( \Tilde{D}^{-\frac{1}{2}} \Tilde{A} \Tilde{D}^{-\frac{1}{2}} H^{l} W^{l} \right),
\label{eq:gcn}
\end{equation}

\noindent where $\sigma$ is an activation function, $\Tilde{A} = A + I$ is the adjacency matrix plus the identity matrix $I$, $\Tilde{D_{ii}} = \sum_j \Tilde{A_{ij}}$, and $W^{l}$ is the parameter matrix of layer $l$. Here, $H^0$ is given by the input matrix $F_M$ and the output matrix $H$ corresponds to the matrix $H^l$ of the last layer. This formulation is especially interesting because it is isomorphic graph invariant, \textit{i.e.}, the result is the same regardless of the order of the nodes (which yields different adjacency and diagonal matrices). Figure~\ref{fig:architecture}(a) shows the schematics of the GCN model. For more implementation details about the GCN, check Section~\ref{subsec:gcn}.

Structure2Vec~\cite{s2vec} is a recent graph embedding approach where each node is codified by considering a given information of the node and the embedding of all of its neighborhood. It is a recursive approach that updates each node embedding through a given number of layers. After $L$ layers, the node embedding $h_i^L$ of node $i$ in layer $L$ considers the topological information of all nodes in its L-hop neighborhood. If we run the method with $d_m$ layers, where $d_m$ is the diameter of the graph, then the node embedding considers the topological information of the entire graph. Since the diameter of real-world networks is usually small compared with the network size, then the number of layers required to embed the whole graph is also small. Formally, the embedding $h_i$ of a node $i$ at layer $l+1$ is defined as a non-linear function $f$ that receives three parameters:

\begin{equation}
h_i^{l+1} = f \left(x_i, \quad \sum_{n\in \mathcal{N}(i)} h_n^{l},  \quad   w(i, n)_{n\in \mathcal{N}}(i)         \right),
\label{eq:embed_simple}
\end{equation}

\noindent where $f$ is any nonlinear function (such as a neural network for example), $\mathcal{N}(i)$ is the set of neighbors of node $i$, $x_i$ is any relevant node information that should be embedded, and $w(i,n)$ is the weight of the edge connecting nodes $i$ and $n$ (in the unweighted setup, such as the one used here, $w(i,n) = 1$ for all existing edges and $w(i,n) = 0$ for non-existing edges). The node embeddings are then summed in order to generate a graph embedding, which is used for graph classification problems. $h_i^0$ is initialized with zeros for all nodes $i$.  Figure~\ref{fig:architecture}(b) shows the Structure2Vec model. More details about the implementation of Structure2Vec are discussed in Section~\ref{subsec:s2vec}.

Hanjun et al.~\cite{combinatorial} adopt a similar approach, also based on the Structure2Vec embedding, but instead of using it for classification problems, the authors adapt the method to work with reinforcement learning in order to solve combinatorial graph problems.

Here, we use the GCN and Structure2Vec methods for our experiments. 

\subsection{Network Centrality Prediction}

There have been several attempts to create reliable methods capable of efficiently predicting node centrality measures in complex networks. A notable method for approximating centrality measures is presented in \cite{sample_between} and \cite{sample_close}, where betweenness and closeness centralities are approximated using a sampling technique, respectively. These methods calculate the single-source shortest path (SSSP) for a given sample of nodes, which is then used to define their exact centrality value. These SSSPs are also used to approximate the centrality value of other non-sampled nodes that belong to the SSSP. Brandes et al.~\cite{sample_parameters} complement these approaches by testing different sampling methods, where they show that the best way to select the node sample is through random selection. Cohen et al.~\cite{aprox_close} improve over the sampling technique for closeness centrality by also using a technique called \textit{pivoting}, which essentially approximates the closeness centrality of a node~$v$ by the value of the closest node $x$ that belongs to the sample set, called a pivot. Other existing sampling methods for approximating the betweenness centrality improved upon the previous ideas by adding an error guarantee value~\cite{fast_bw} and adapting sampling techniques for evolving graphs~\cite{aprox_evolve_1}. However, as stated by Grando et al.~\cite{grando_3}, sampling techniques are still computationally expensive, given that even calculating the exact betweenness or closeness centrality values for only a small portion of the nodes is still costly.

More recently, Grando et al.~\cite{grando_1} presented a more general approach that approximates 8 different node centralities by using a neural network (NN). The NN model predicts one centrality value by receiving as inputs for each node the 7 remaining centrality values. The main novelty of this approach is that it is very fast to predict a node centrality after the NN is properly trained. The drawback is that this model requires the computation of 7 node centralities in order to predict the 8th centrality value for each node. The authors then expanded upon their initial idea~\cite{grando_2,grando_3} by creating a NN model that receives only 2 centrality values as an input for each node in order to predict any other centrality: (i) degree and (ii) eigenvector centralities. This marks a great improvement over the initial idea, given that the degree and eigenvector centralities both have a low computational cost and can be used to predict the values of centralities with a high computational cost, such as betweenness and closeness. The NN is trained using a set of artificial complex networks with different sizes~(varying from 100 to 1000 nodes), and the exact centrality measures for each node. After the model is trained, it generalizes well for other networks, even larger ones.

We present here a similar approach for predicting node centralities that requires less information for each node and that achieves better results than the work presented by Grando et al.~\cite{grando_3}. In Section~\ref{sec:method}, we present the details of our model and how it differentiates from the current state-of-the-art.


\section{NCA-GE Model}
\label{sec:method}

We present here the Network Centrality Approximation using Graph Embedding~(NCA-GE) model, capable of approximating any node centrality using only the degree centrality as feature input. The approach outlined in this paper is inspired by two distinct approaches: (i)~the NN node centrality prediction model recently introduced by Grando et al.~\cite{grando_3}; and (ii)~modern graph embedding techniques, such as the Graph Convolutional Network (GCN)~\cite{gcn} and Structure2Vec~\cite{s2vec}. The proposed idea is to codify each node into an embedding vector calculated using only the degree centrality as a feature of a given node. We then pass these embeddings to a NN~model~(Figure~\ref{fig:architecture}(c)), which outputs an approximation of the desired centrality. The output centrality can be any centrality other than the degree (since it is already used as input). The NCA-GE can use any graph embedding technique, but here we explore using two specific methods: GCN and Structure2Vec. Our contributions are three-fold: (i) our approach achieves better results for artificial and real-world networks when compared with previous work~\cite{grando_3}, (ii) the NCA-GE uses less information as input, making it easier to be executed, as we show in Section~\ref{subsec:time}, and (iii) our model is more customizable in the sense that it works with any graph embedding method, allowing it to be optimized for different scenarios (different network topologies, for example) using different graph embedding methods.

In the remainder of this section, we discuss our method in more detail. We present the implementation details for the GCN and Structure2Vec embedding methods used. The training process is also discussed. We also present the implementation details of the baseline model used to compare with our approach. The training parameters used for each of the models presented here are shown in Table~\ref{tab:param}.

\begin{table*}[ht]
	\centering
	\small{
	\caption{Parameters used for training the proposed model and the baseline model. \textit{GE Layers} refers to the number of iterations are used to compute the embedding matrix (Figure~\ref{fig:architecture}(a) and \ref{fig:architecture}(b)); $\eta$ is the learning rate used; $\beta$ is the learning rate decay; $\lambda$ refers to the weight given to the regularization term in the loss function; $C$ is the number of input features used; and $F$ is the size of the embedding vectors generated.}
	\label{tab:param}
	\begin{tabular}{|l|c|c|c|c|c|c|c|c|}
		\hline
		& \textbf{GE Layers} & \textbf{\boldmath{$\eta$}}  & \textbf{\boldmath{$\beta$}}  & \textbf{min $\eta$}  & \textbf{\boldmath{$\lambda$}} & \textbf{C} & \textbf{F} & \textbf{Batch Size}	\\ \hline
		\textbf{NCA-GE (GCN)}	& 2                 & 0.001         & 0.999             & 0.0001	        & 0.01	            & 1          & 128          & 128       \\ \hline
		\textbf{NCA-GE (S2VEC)} & 2                 & 0.001         & 0.999             & 0.0001            & 0.1 	            & 1			 & 128          & 128       \\ \hline
		\textbf{Baseline} 	    & -                 & 0.01          & 0.999             & 0.0001	        & 0.001	            & 2		     & 128          & 128       \\ \hline
	\end{tabular}%
}
\end{table*}

\subsection{GCN: Implementation Details}
\label{subsec:gcn}

Here we detail how we implemented the GCN method. A schematic of this model is presented in Figure~\ref{fig:architecture}(a). The feature matrix $F_M$ (Eq.~\ref{eq:gcn}) used here is comprised solely of the rank of the degree centrality of the nodes in the graph. The nodes are embedded through a recursive process, as depicted in Eq.~\ref{eq:gcn}, where the number of layers determines the number of iterations. We parameterize the node embedding process of Eq.~\ref{eq:embed_simple} as follows:

\begin{equation}
H^{l+1} = relu \left( \Tilde{D}^{-\frac{1}{2}} \Tilde{A} \Tilde{D}^{-\frac{1}{2}} H^{l} W^{l} \right),
\label{eq:embed_parameter}
\end{equation}

\noindent where $relu(.)$ is the rectified non-linearity function and the remaining variables are the same as in Eq.~\ref{eq:gcn}. We used 2 layers, where $W^1 \in \rm I\!R^{CxF}$, $W^2 \in \rm I\!R^{FxF}$, $C$ being the number of initial features and $F$ being the embedding size. As already mentioned, here we use only a single feature: the normalized degree rank of a node, thus, $C = 1$.

\subsection{Structure2Vec: Implementation Details}
\label{subsec:s2vec}

We implemented the Structure2Vec model following its definition in Eq.~\ref{eq:embed_simple}. The Structure2Vec model is presented in Figure~\ref{fig:architecture}(b). Eq.~\ref{eq:embed_simple} only defines the embedding of single nodes, but for an implementation convenience, we compute the embedding $h_i^l$ of all nodes $i \in V$ at layer $l$ in a single matrix operation, and store each embedding in the rows of matrix $H^l$ (Figure~\ref{fig:architecture}(b)). Similarly to Hanjun et al.~\cite{combinatorial}, we also added a weight matrix in the embedding loop to account for the connection weights of the graph, represented in Figure~\ref{fig:architecture}(b) by the upper flow. The model is parameterized as follows:

\begin{eqnarray}
d_i &=& \sum_{j=1}^{N} A_{ij}, \forall i = 1, ..., N \\
X_1 &=&  d W^1 \\
X_3 &=& F_M W^3 \\
H^0 &=& relu(X_1 + X_3) \\
H^{l+1} &=& relu \left( A H^l W^2 + X_1 + X_3 \right),
\label{eq:s2vec_parameter}
\end{eqnarray}

\noindent where $d \in \rm I\!R^{Nx1}$ is the degree vector, $W^1, W^2 \in \rm I\!R^{FxF}$ and $W^3 \in \rm I\!R^{CxF}$ are the weight matrices, and $H^l$ is the embedding matrix at layer $l$. We used 2 layers, therefore, the final embedding matrix is given by $H^2$. Notice that Structure2Vec uses the plain adjacency matrix $A$, unlike the GCN, which uses a normalized version of $A$ with self-loops.

\subsection{NCA-GE}

After computing the embedding matrix $H$ (using whichever graph embedding method), we select only the embeddings of the nodes being analyzed in the current batch. The selected nodes are given by:

\begin{equation}
H_D = I_D \times H,
\label{eq:mu}
\end{equation}

\noindent
where $I_d \in \rm I\!R^{BxN}$ is the one-hot encoding of the id of the nodes in the batch and $B$ represents the size of the batch. The centrality prediction outputted by the NN is then given by:

\begin{equation}
Y = relu(relu(relu(H_D W^{F1}) W^{F2}) W^{F3}) W^{F4},
\label{eq:NN}
\end{equation}

\noindent where $W^{F1}, W^{F2}, W^{F3}, W^{F4} \in \rm I\!R^{FxF}$ are the weight matrices. 

As we can see from Figure~\ref{fig:architecture}, all weight matrices used are independent of the graph size $N$. This allows our approach to be used to predict node centralities for graphs of any size. We can also notice that both graph embedding models used here (GCN and Structure2Vec) require the complete adjacency matrix $A$ and the one-hot encoding of the nodes in the current batch. To achieve a fast training time with our model, we use sparse tensor representations implemented in Tensorflow.

The feature input of the NCA-GE is the normalized rank value of the degree centrality of a given node and the output is the normalized value of the rank of the desired higher-order centrality, both within the range [0,~1]. Therefore, the only feature information required by our model is the degree centrality of each node, which is a very low cost and simple centrality to be obtained. We experienced using the normalized value of the exact degree centrality as an input, but the results were not as good as using the normalized value of the degree rank, even when we intended to predict the normalized value of the exact centrality as output. 

The error function used for our NN model is the mean squared error (MSE) between the predicted centrality and the expected value. We also used an L2 regularization function to avoid over-fitting. Therefore, the adopted loss function is:

\begin{equation}
\mathcal{L}_{TOTAL} = \mathcal{L}_{MSE} + \lambda \mathcal{L}_{REGUL},
\label{eq:loss}
\end{equation}

\noindent where $\mathcal{L}_{MSE}$ is the loss function for the Mean Squared Error, $\mathcal{L}_{REGUL}$ is the L2 regularization loss, and $\lambda$ is a variable that controls the weight given to the regularization term. Regularization loss helps to avoid the model to become too overfitted to the training data, losing its generalization capabilities. We used the Adam algorithm as the optimization function, the gradients were truncated in the range [-1, 1], and the batch size was set to 128.

\subsection{Training Procedure}
\label{sec:training}

The training procedure adopted is similar to the methodology used by Grando et al.~\cite{grando_3}: We train the NN over a set of 1000 artificial complex networks, with the network size ranging from 100 to 1000 nodes. All networks were created using the scale-free generation model~\cite{barabasi}. Each batch of data passed to the model contains data of a single graph. This allows us to pass a single adjacency matrix for each batch.

It is important to note that the model is trained separately for each centrality. Therefore, for instance, when the model is trained using the betweenness centrality, it will not be able to accurately predict other centralities. This is not an issue, however, given that the NCA-GE trains quickly, as it uses only small synthetic graphs during training.

During training, we decay the learning rate after each batch is processed. We decay our learning rate according to:

\begin{equation}
\eta = \eta \beta,
\end{equation}

\noindent where $\eta$ is the learning rate and $\beta$ refers to the learning rate decay coefficient. The learning rate is decayed until it reaches a minimum value. Check Table~\ref{tab:param} for a full list of the parameters used. Finally, the gradients used to update the weight matrices are clipped in the range of [-1,~1] in order to avoid instabilities during training.

\subsection{Baseline Model: Implementation Details}
\label{subsec:baseline}

We reproduced the model created by Grando et al.~\cite{grando_3} in order to compare their results with ours, given that this is to the best of our knowledge the most similar work to ours in the literature, while also being a very recent study. This model will be called the \emph{baseline} model from now on. In the remainder of this section, we detail our implementation of the baseline model. 

The model proposed by Grando et al.~\cite{grando_3} receives the degree and eigenvector centralities as input and outputs the desired higher-order centrality as output (such as betweenness, for example). The main advantage of the NCA-GE over the baseline model is that our approach uses only the degree centrality as a feature input, not requiring the eigenvector centrality. We follow the neural network architecture described in the original paper~\cite{grando_3}: The inputs are the normalized ranks of degree and eigenvector centralities in the range [-1,~1], and the output is the normalized rank of the desired node centrality, also in the range [-1,~1] (note that this differs from our approach, where the input and output are in the range of [0,~1]). The model consists of four fully connected hidden layers with 20 neurons each, where each hidden layer is followed by the hyperbolic tangent activation function. We used the same loss function that we used for our model, depicted in Equation~\ref{eq:loss} (the original paper did not use a regularization term, but this showed to output better results during preliminary experiments). We used the Adam optimizer as the optimization function, differently from the Levenberg-Marquardt optimizer used in the original paper. The gradients were also clipped in the range of [-1,~1] similarly to our approach, but this process is not specified in the original paper.

\section{Performance Evaluation}
\label{sec:result}

\begin{figure*}[ht]
	\centering
	\subfigure[Betweenness Rank.]{\includegraphics[scale=0.47]{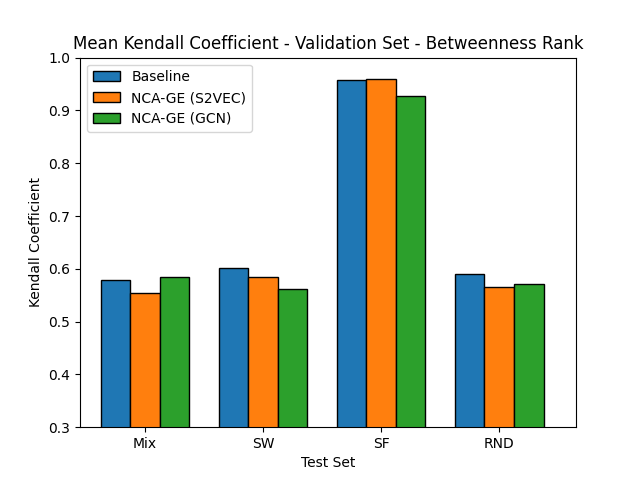}}
	\subfigure[Closeness Rank.]{\includegraphics[scale=0.47]{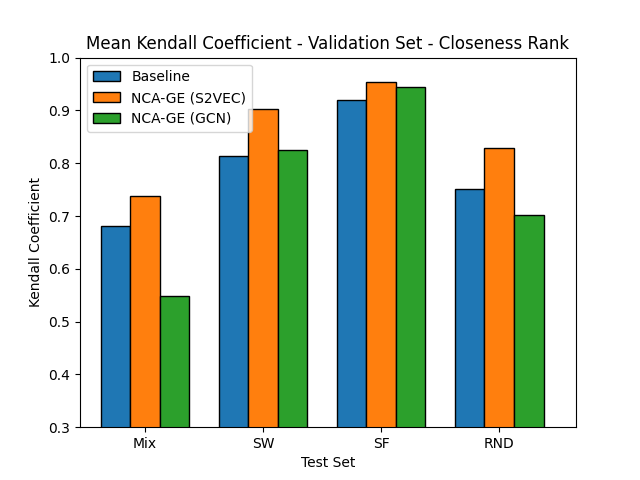}}
	\subfigure[Harmonic Rank.]{\includegraphics[scale=0.47]{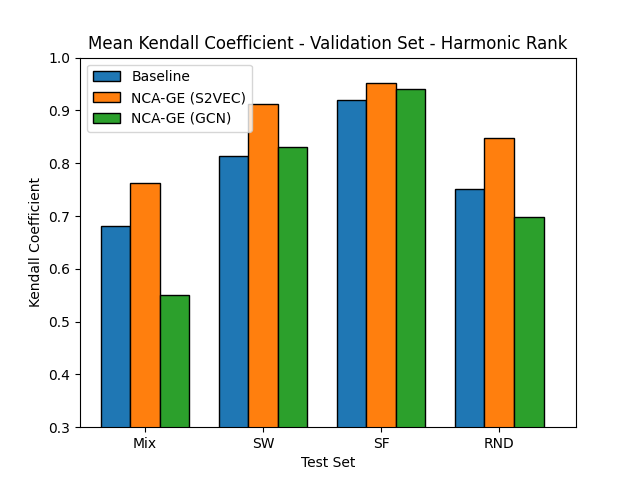}}
	\subfigure[Eigenvector Rank.]{\includegraphics[scale=0.47]{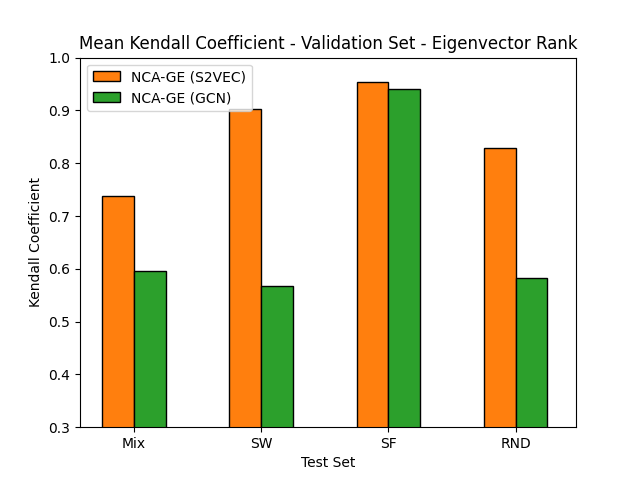}}
	
	\caption{Mean Kendall coefficient for the synthetic network for the baseline, NCA-GE (S2VEC), and NCA-GE (GCN) models. Each result is the average of 5 executions. We separated the results based on 4 different test sets: small-world set (SW), scale-free set (SF), random graphs set (RND), and a mixed set containing all previous 3 graph types (MIX). Note that we do not compare the results for the Eigenvector centrality with the baseline model since it can not predict this centrality, given that this centrality is one of its inputs. The standard deviation was not included in the graph as it presented very small values for all three models presented (values ranging from $10^{-3}$ to $10^{-5}$).}
	\label{fig:kendall}
\end{figure*}

\begin{figure*}[ht]
	\centering
	\subfigure[Betweenness Rank.]{\includegraphics[scale=0.47]{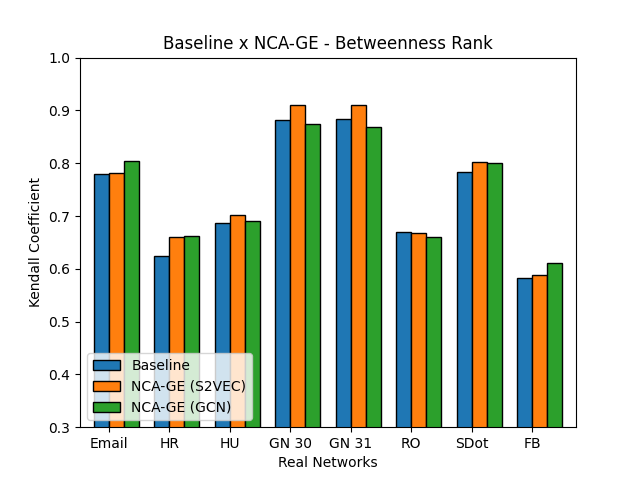}}
	\subfigure[Closeness Rank.]{\includegraphics[scale=0.47]{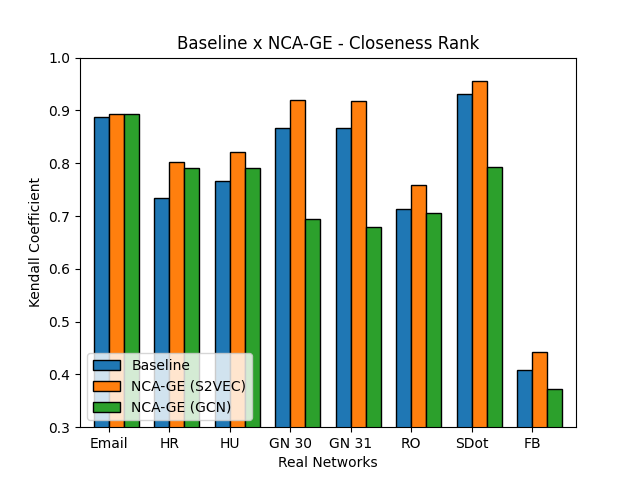}}
	\subfigure[Harmonic Rank.]{\includegraphics[scale=0.47]{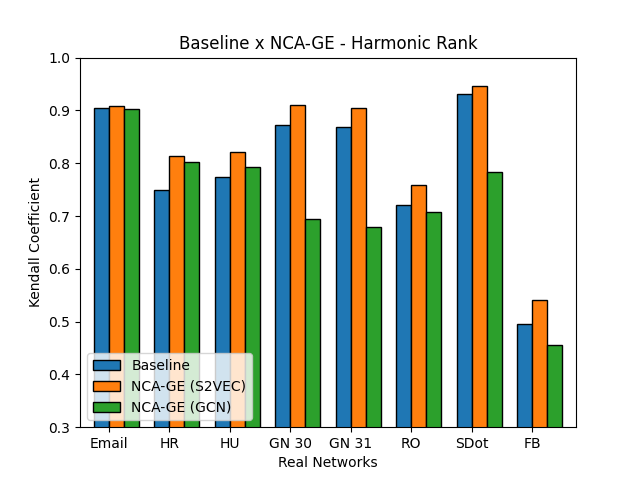}}
	\subfigure[Eigenvector Rank.]{\includegraphics[scale=0.47]{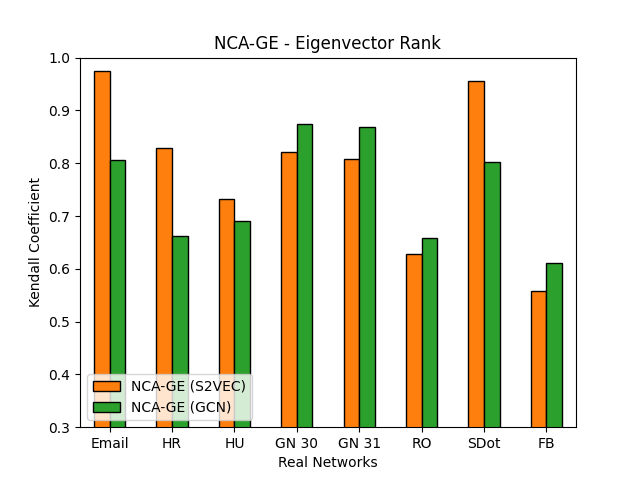}}
	
	\caption{Comparison of the mean Kendall Rank Coefficient for the real-world networks described in Table~\ref{tab:networks} when using the baseline,  NCA-GE (S2VEC), and NCA-GE (GCN). These results are computed using an average of 5 executions. Note that we do not compare the results for the Eigenvector centrality with the baseline model since it can not predict this centrality, given that this centrality is one of its inputs. The standard deviation was not included in the graph as it presented very small values for all three models presented (values ranging from $10^{-3}$ to $10^{-5}$).}
	\label{fig:kendall_real}
\end{figure*}

\begin{figure}[ht]
	\centering
	\includegraphics[scale=0.75]{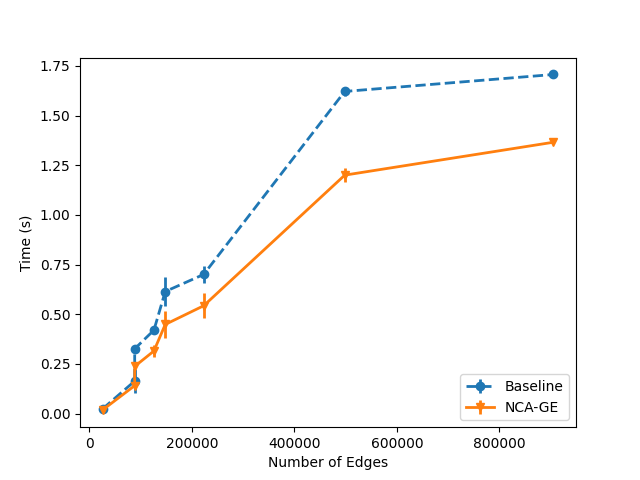}
	
	\caption{Comparison of the execution time for the real graphs using the baseline and NCA-GE models in relation to the number of edges in the network. The Y-axis represents the time required to process the centrality for all nodes in a graph. The X-axis represents the number of edges in the network. These values represent the average of 5 executions. The error bars for each point represent the standard deviation.}
	\label{fig:time}
\end{figure}

As detailed in Section~\ref{sec:training}, we train the proposed NCA-GE model by iterating over all nodes in a set of training graphs. Each node fed to the model is considered a step. The results presented here in this section for the NCA-GE model were all generated after training for 4 million steps, while the baseline model was trained for 1 million steps.

As stated in Grando et al.~\cite{grando_3}, the MSE is not a reliable metric to evaluate how well a model that predicts the centrality rank of a node is performing. This is due to the fact that we are interested in a model that predicts the correct rank ordering of a given centrality measure for all nodes, not the resulting absolute rank value. Therefore, we follow the methodology adopted by Grando et al.~\cite{grando_3} and use the Kendall $\tau$-b~\cite{kendall} rank correlation coefficient to determine the quality of the NCA-GE rank prediction for network centralities. The Kendall $\tau$-b coefficient varies in the range of [-1,~1] and it measures the correlation between two ranking lists. A value of -1 indicates that the two lists are fully reversed (that is, if index X is ranked first in one list, it is ranked last in the other, and so on for the other rank positions). A value of~0 means that there is no correlation, whereas a value of~1 indicates that both lists fully agree with the ranks of every node. Therefore, to measure the performance of a prediction model, we first build two rank lists: (i)~a list in which the value in index $i$ indicates the true normalized rank of node~$i$, and (ii)~a similar list that stores the predicted normalized rank. We then measure the Kendall $\tau$-b coefficient between these two lists, where a value near 1 indicates good performance since the prediction model managed to rank most of the nodes correctly.

We tested the NCA-GE model with four different node centralities: (i) betweenness, (ii) closeness, (iii) harmonic, and (iv) eigenvector. The first three centralities are very computationally expensive, as discussed in Section \ref{sec:basic}, while the fourth centrality is relatively cheap, but we experimented with it in order to show the effectiveness of the NCA-GE model. 

In this section, we detail several experiments that we performed to showcase the efficiency of the proposed NCA-GE approach using the GCN or Structure2Vec graph embedding methods. We use as a baseline to compare our results the model recently proposed by Grando et al.~\cite{grando_3}~(see Section~\ref{subsec:baseline}). We evaluate and compare our approach in a series of artificial complex networks generated by different generation models and a set of real-world networks. We also analyze the time complexity of both baseline and NCA-GE models. All experiments shown here for the NCA-GE model were generated using only the normalized rank degree centrality as feature input for each node (as previously stated in Section~\ref{sec:method}), while the results for the baseline model were generated using the normalized rank degree and eigenvector centralities as input. These experiments aim to show that the NCA-GE achieves comparable or better results when compared with the baseline model, and it manages this by using only the degree centrality as a feature input.

\subsection{Experimental Results with Synthetic Networks}

For our first experiment, we use the NCA-GE model to predict the normalized centrality rank of all nodes of a set of synthetically generated networks. To that end, we created 4 test network sets, each one built using different generation models: (i)~random networks~\cite{erdos}; (ii)~small-world~\cite{Watts1998Collective}; (iii)~scale-free~\cite{barabasi}; and (iv)~a mixture of all three types of networks. Each set contains 100 artificial graphs. Our aim with this experiment is to show how the NCA-GE and the baseline models behave for networks with different topologies. As previously mentioned, for each analyzed network, we measure the Kendall $\tau$-b coefficient between the true rank of each node and the predicted ranks. Hence, each network is associated with a Kendall $\tau$-b coefficient indicating how well the rank for a given centrality measure is predicted.

Figure~\ref{fig:kendall} shows the mean Kendall $\tau$-b coefficient for the networks of each of the test sets and for the four previously mentioned centralities using the baseline model, the NCA-GE using GCN (NCA-GE (GCN)), and NCA-GE using Structure2Vec (NCA-GE (S2VEC)). These results were obtained by executing each experiment five times and then taking the mean. The standard deviation was not included in the graph as it presented very small values for all three studied models  (values ranging from $10^{-3}$ to $10^{-5}$). We were not able to compare the NCA-GE with the baseline model for the eigenvector centrality, since the latter model uses this centrality as its input. 

From Figure~\ref{fig:kendall}, we observe that all three methods performed very well for all centralities for the artificial scale-free networks, which makes sense since the training set used for all experiments was comprised entirely of artificial scale-free networks. We can also see that the NCA-GE (S2VEC) performs as well as the baseline model for the betweenness centrality while presenting a better result for the closeness and harmonic centralities when compared to the baseline for all types of network topologies. This shows just how powerful the NCA-GE with Structure2Vec is: It manages to achieve equal to better results when compared with the baseline model without using any higher-order feature, such as the eigenvector centrality. Regarding the NCA-GE using the GCN embedding, we can note that it did not excel as its Structure2Vec counterpart, although it still managed to achieve comparable results to the baseline model for the closeness and harmonic centralities, and almost equal results for betweenness. The results for the eigenvector centrality evidences the superiority of the NCA-GE using Structure2Vec over the GCN for computing the node embeddings.

As a final remark, in Figure~\ref{fig:kendall}, we can observe that the results for closeness and harmonic centralities are very similar. This phenomenon can be explained by the similarities in how closeness and harmonic centralities are computed. We can also note that the lowest obtained Kendall coefficients are for the betweenness centrality. This is likely due to the fact that closeness and harmonic centralities are more stable across different topologies, since this centrality tends to have low values and low deviation. On the other hand, the distribution of values for the betweenness centrality tends to variate more depending on the graph topology. This can be noted by the high Kendall coefficients for all models for the SF topology, which was also used during training, but low values for the remaining topologies.

\subsection{Experimental Results with Real-World Networks}

We now evaluate the NCA-GE model with real-world networks. These networks were extracted from the \textit{Stanford Large Network Dataset Collection}~\cite{snap}. The networks we use and their respective properties are listed in Table~\ref{tab:networks}. The ``Density" property is simply the density of the adjacency matrix of a graph, given by the number of actual edges divided by the total number of possible edges in a graph.

\begin{table*}[ht]
	\centering
	\footnotesize{
	\caption{Statistics for the real-world networks used in the experiments.}
	\label{tab:networks}
	\begin{tabular}{|l|c|c|c|c|c|c|}
		\hline
		& \textbf{Abbreviation} & \textbf{Nodes} & \textbf{Edges} 	& \textbf{Avg. Degree}  & \textbf{Avg. Clustering Coef.} & \textbf{Density}	\\ \hline
		\textbf{email-Eu-core}	    & Email      & 1,005    & 25,571   & 33.246	 & 0.399	 & 0.025			\\ \hline
		\textbf{facebook\_combined} & FB         & 4,039    & 88,234   & 43.691  & 0.606 	 & 0.005			\\ \hline
		\textbf{Deezer\_RO} 	    & RO         & 41,773   & 125,826  & 6.024	 & 0.091	 & 7.211e-05		\\ \hline
		\textbf{Deezer\_HR} 	    & HR         & 54,573   & 498,202  & 18.26	 & 0.136	 & 1.673e-04		\\ \hline
		\textbf{Deezer\_HU} 	    & HU         & 47,538   & 222,887  & 9.377	 & 0.116	 & 1.012e-05		\\ \hline
		\textbf{p2p-Gnutella30} 	& GN 30      & 36,682   & 88,328   & 4.816	 & 0.006	 & 6.564e-05		\\ \hline
		\textbf{p2p-Gnutella31} 	& GN 31      & 62,586   & 147,892  & 4.726	 & 0.005	 & 3.776e-05		\\ \hline
		\textbf{soc-Slashdot0811} 	& SDot       & 77,360   & 905,468  & 14.13	 & 0.055	 & 1.513e-04		\\ \hline
	\end{tabular}%
	}
\end{table*}

Figure \ref{fig:kendall_real} presents the results of the average Kendall $\tau$-b coefficient of all four centralities for the real networks~(Table~\ref{tab:networks}) using the baseline, the NCA-GE~(GCN), and NCA-GE~(S2VEC) models. The obtained results for the real networks are somewhat similar to the obtained results for the synthetic test sets: The NCA-GE using Structure2Vec managed to achieve equal to better results for all networks and all centralities (except eigenvector centrality) when compared with the baseline model. Once again, these results were obtained using only the normalized degree rank as input feature, contrasting to the baseline model, which uses the degree and eigenvector centralities as input features. The results for NCA-GE using GCN were also similar to the ones obtained for the synthetic test sets, where it achieved similar results to the baseline model for the betweenness centrality, and slightly worse to equal results for the closeness and harmonic centralities. For the eigenvector centrality, we can see a better performance of the NCA-GE with GCN when compared with the obtained results for the test sets: Here it manages to achieve slightly better results than the NCA-GE with Structure2Vec for some graphs, although the latter still outperforms the former in the majority of the real-world graphs.

\subsection{Time Complexity Analysis}
\label{subsec:time}

Predicting node centrality is only useful if the prediction model is more time-efficient than calculating the true centrality values. To assess this topic, we measured the processing time required for computing the betweenness centrality rank for all nodes in the real-world networks presented in Table~\ref{tab:networks}. We used only one of the four centralities tested because the time required to compute such values does not depend on the target centrality. Therefore, the time measurements obtained for the betweenness centrality applies to the other three centralities as well. We measured the processing time for the baseline and the NCA-GE with Structure2Vec. The complexity of the NCA-GE using GCN is the same as using the Structure2Vec. All experiments were performed on a machine with the following configurations: Ubuntu 18.04, Intel Core i7-7700K octa-core CPU with 4.20GHz, Nvidia GTX 1080 with 8Gb of internal memory, 32Gb of RAM, using Tensorflow 1.14 with GPU support. We only measure the processing time of the model and the computation of the input features. Therefore, for the baseline model, we take also into account the time for computing the eigenvector centrality, given as input to this model. Similarly, for the NCA-GE, we take also into account the time required to fetch the sparse representation of the adjacency matrix, given as input to the model.

The baseline model requires the degree and eigenvector centralities as input for the model. Considering a graph $G=(V,E)$, where $V$ is the set of vertices (nodes) and $E$ is the set of edges in the graph, to obtain the degree of all nodes has a time complexity of $O(N)$, where $N=|V|$. The eigenvector centrality is formally known to have a time complexity of $O(N^2)$, but using the power method~\cite{powermethod}, this complexity falls to $O(k|E|)$, where $k$ is the number of iterations required by the power method to converge, which is difficult to predict beforehand. After obtaining these two centralities for the input, the baseline model performs a constant number of operations for each node in order to compute its target centrality rank, resulting in a total time complexity of $O(N + k|E| + c_b N) = O(k|E|)$, where $c_b$ is a constant reflecting the number of operations performed by the baseline model. 

\begin{table}[ht]
	\centering
	\small{
	\caption{Mean execution time (s) and standard deviation for processing the betweenness centrality using the baseline and NCA-GE models for each of the real networks used for validation. These values represent the average of 5 executions.}
	\label{tab:time}
	\begin{tabular}{|l|c|c|c|}
		\hline
		\textbf{Graph} & \textbf{Baseline} & \textbf{NCA-GE}       \\ \hline
		\textbf{Email}  & 0.024 $\pm$ 0.001   & 0.018 $\pm$ 0.001  \\ \hline
		\textbf{FB} 	& 0.164 $\pm$ 0.007   & 0.141 $\pm$ 0.002  \\ \hline
		\textbf{RO} 	& 0.420 $\pm$ 0.022   & 0.315 $\pm$ 0.033  \\ \hline
		\textbf{HR} 	& 1.621 $\pm$ 0.074   & 1.200 $\pm$ 0.067  \\ \hline
		\textbf{HU} 	& 0.701 $\pm$ 0.016   & 0.543 $\pm$ 0.008  \\ \hline
		\textbf{GN 30} 	& 0.326 $\pm$ 0.011   & 0.238 $\pm$ 0.029  \\ \hline
		\textbf{GN 31} 	& 0.615 $\pm$ 0.060   & 0.449 $\pm$ 0.006  \\ \hline
		\textbf{SDot} 	& 1.707 $\pm$ 0.042   & 1.366 $\pm$ 0.063  \\ \hline
	\end{tabular}%
	}
\end{table}

The NCA-GE model requires the degree centrality and the sparse representation of the adjacency matrix as input, which has a time complexity of $O(N)$ and $O(|E|)$, respectively. After that, the model performs a series of matrix multiplications, where these operations are bounded by the matrix operations using the adjacency matrix $A$, with dimensions $N \times N$, resulting in a time complexity of $O(N^2)$. However, since we use sparse matrix representations, this complexity becomes associated with the density of the matrix, i.e, $O(|E|)$. Note that the NCA-GE computes the target centrality of all nodes in a single pass. Therefore, the overall complexity of the NCA-GE model is given by $O(N + |E| + c_p|E|) = O(|E|)$, where $c_p$ is a constant reflecting the number of operations performed by the proposed NCA-GE model.  

Figure \ref{fig:time} presents the time required to process the centrality rank of every node in relation to the number of edges for the real-world networks using the baseline and NCA-GE with Structure2Vec. Table~\ref{tab:time} presents the execution time for each of the networks we use. As we can see, the time required to process a network of both models grows linearly with the number of edges, confirming our time complexity analysis of $O(|E|)$. The processing time gap between the baseline model and the NCA-GE can be explained by the former's time complexity ($O(k|E|)$) being greater than the latter ($O(|E|)$). Since this difference is given only by the number of iterations $k$ of the power method, this value still maintains the behavior of $O(|E|)$. This makes the processing time of the baseline model to follow the same behavior of the NCA-GE, albeit with a slight shift in the time axis, as depicted in Figure~\ref{fig:time}.

\section{Discussion and Final Remarks}
\label{sec:conclusion}

We presented the Network Centrality Approximation using Graph Embedding~(NCA-GE) model, which enables approximating node centralities of large real-world networks using only node degrees as a feature input. NCA-GE uses graph embedding techniques to build a powerful internal node representation of each node, which is then used to predict the rank of the desired centrality. NCA-GE is independent of the adopted graph embedding technique.
Moreover, NCA-GE is trained using only small artificial scale-free graphs, which shows that the proposed approach is robust.

NCA-GE was compared with the current state-of-the-art model for predicting node centrality based on Machine Learning~\cite{grando_3}, achieving an equal to better performance for a set of real-world networks. Nevertheless, the main advantage of the NCA-GE over the baseline model is that the latter uses the degree and the eigenvector centrality of each node in order to predict the target centrality, whereas the former uses only the degree information as a feature input. We also showed how the choice of the graph embedding method used by the NCA-GE can impact the results. We also showed that the proposed NCA-GE model is very time efficient, being able to compute the desired centrality of all nodes in a large network within a very small timeframe. 

As previously stated, here we only consider unweighted and undirected graphs. However, it would be interesting to see how the NCA-GE handles directed and weighted graphs. Dealing with weighted graphs is more straightforward, since we only need to consider a weighted adjacency matrix $A$ (or its normalized version $\Tilde{D}^{-\frac{1}{2}} \Tilde{A} \Tilde{D}^{-\frac{1}{2}}$) and the weighted degree centrality (the degree $d_i$ of node $i$ is given by the sum of weights of all edges connected to $i$). Considering a directed graph is more complicated, given that the graph embedding method used must be able to deal with directed edges (which is not the case for the Structure2Vec and GCN).

As future work, we intend to expand NCA-GE to predict other node centrality measures
or
other node-specific features. We aim to adapt the NCA-GE to predict a node's time centrality~\cite{time_cent} in relation to its diffusive capabilities in the network by 
embedding nodes in Time-Varying Graphs~(TVGs)~\cite{tvg_embed}, allowing the NCA-GE to predict diffusion-related centralities~\cite{new_rank,time_evol}. Another possible future work is to adapt the NCA-GE to work with directed graphs using other graph embedding approaches, as well as to test the NCA-GE with weighted graphs. Finally, given the difference in results obtained by the NCA-GE using varying graph embedding methods (GCN and Structure2Vec), we aim to investigate and better understand 
NCA-GE with other graph embedding methods and see how each one performs for different network topologies.

\section*{Acknowledgment}

This work has been partially supported by CAPES, CNPq, FAPERJ, and FAPESP. Authors also acknowledge the INCT in Data Science -- INCT-CiD.

\bibliographystyle{IEEEtran}
\bibliography{Ref}

\begin{thebibliography}{10}
\providecommand{\url}[1]{#1}
\csname url@samestyle\endcsname
\providecommand{\newblock}{\relax}
\providecommand{\bibinfo}[2]{#2}
\providecommand{\BIBentrySTDinterwordspacing}{\spaceskip=0pt\relax}
\providecommand{\BIBentryALTinterwordstretchfactor}{4}
\providecommand{\BIBentryALTinterwordspacing}{\spaceskip=\fontdimen2\font plus
\BIBentryALTinterwordstretchfactor\fontdimen3\font minus
  \fontdimen4\font\relax}
\providecommand{\BIBforeignlanguage}[2]{{%
\expandafter\ifx\csname l@#1\endcsname\relax
\typeout{** WARNING: IEEEtran.bst: No hyphenation pattern has been}%
\typeout{** loaded for the language `#1'. Using the pattern for}%
\typeout{** the default language instead.}%
\else
\language=\csname l@#1\endcsname
\fi
#2}}
\providecommand{\BIBdecl}{\relax}
\BIBdecl

\bibitem{app_cn_3}
L.~Maccari and R.~L. Cigno, ``Pop-routing: Centrality-based tuning of control
  messages for faster route convergence,'' in \emph{IEEE INFOCOM 2016 - The
  35th Annual IEEE International Conference on Computer Communications}, 2016,
  pp. 1--9.

\bibitem{malha_aerea}
K.~Wehmuth, B.~B.~A. Costa, J.~V.~M. Bechara, and A.~Ziviani, ``A multilayer
  and time-varying structural analysis of the brazilian air transportation
  network,'' in \emph{Proceedings of the Latin America Data Science Workshop},
  2018, pp. 57--64.

\bibitem{app_b_2}
M.~Rubinov and O.~Sporns, ``Complex network measures of brain connectivity:
  Uses and interpretations,'' \emph{NeuroImage}, vol.~52, no.~3, pp. 1059 --
  1069, 2010, computational Models of the Brain.

\bibitem{pinsage}
\BIBentryALTinterwordspacing
R.~Ying, R.~He, K.~Chen, P.~Eksombatchai, W.~L. Hamilton, and J.~Leskovec,
  ``Graph convolutional neural networks for web-scale recommender systems,'' in
  \emph{Proceedings of the 24th ACM SIGKDD International Conference on
  Knowledge Discovery \&\#38; Data Mining}, ser. KDD '18.\hskip 1em plus 0.5em
  minus 0.4em\relax New York, NY, USA: ACM, 2018, pp. 974--983. [Online].
  Available: \url{http://doi.acm.org/10.1145/3219819.3219890}
\BIBentrySTDinterwordspacing

\bibitem{graph_rl_survey}
M.~R.~F. Mendon\c{c}a, A.~Ziviani, and A.~M.~S. Barreto, ``Graph-based skill
  acquisition for reinforcement learning,'' \emph{ACM Computing Surveys},
  vol.~52, no.~1, pp. 6:1--6:26, 2019.

\bibitem{app_sn_2}
J.~A. Danowski and N.~T. Cepela, ``Automatic mapping of social networks of
  actors from text corpora: Time series analysis,'' in \emph{2009 International
  Conference on Advances in Social Network Analysis and Mining}, 2009, pp.
  137--142.

\bibitem{grando_3}
F.~Grando, L.~Z. Granville, and L.~C. Lamb, ``Machine learning in network
  centrality measures: Tutorial and outlook,'' \emph{ACM Computing Surveys},
  vol.~51, no.~5, pp. 102:1--102:32, Oct. 2018.

\bibitem{survey_social_net}
K.~Das, S.~Samanta, and M.~Pal, ``Study on centrality measures in social
  networks: a survey,'' \emph{Social Network Analysis and Mining}, vol.~8,
  no.~1, p.~13, Feb 2018.

\bibitem{sample_between}
\BIBentryALTinterwordspacing
D.~A. Bader, S.~Kintali, K.~Madduri, and M.~Mihail, ``Approximating betweenness
  centrality,'' in \emph{Proceedings of the 5th International Conference on
  Algorithms and Models for the Web-graph}, ser. WAW'07.\hskip 1em plus 0.5em
  minus 0.4em\relax Berlin, Heidelberg: Springer-Verlag, 2007, pp. 124--137.
  [Online]. Available: \url{http://dl.acm.org/citation.cfm?id=1777879.1777889}
\BIBentrySTDinterwordspacing

\bibitem{sample_close}
D.~Eppstein and J.~Wang, ``Fast approximation of centrality,'' in
  \emph{Proceedings of the Twelfth Annual ACM-SIAM Symposium on Discrete
  Algorithms}, ser. SODA '01.\hskip 1em plus 0.5em minus 0.4em\relax Society
  for Industrial and Applied Mathematics, 2001, pp. 228--229.

\bibitem{sample_parameters}
U.~Brandes and C.~Pich, ``Centrality estimation in large networks,''
  \emph{International Journal of Bifurcation and Chaos}, vol.~17, no.~07, pp.
  2303--2318, 2007.

\bibitem{aprox_close}
E.~Cohen, D.~Delling, T.~Pajor, and R.~F. Werneck, ``Computing classic
  closeness centrality, at scale,'' in \emph{Proceedings of the Second ACM
  Conference on Online Social Networks}, ser. COSN '14.\hskip 1em plus 0.5em
  minus 0.4em\relax ACM, 2014, pp. 37--50.

\bibitem{parallel_cent}
D.~A. Bader and K.~Madduri, ``Parallel algorithms for evaluating centrality
  indices in real-world networks,'' in \emph{2006 International Conference on
  Parallel Processing (ICPP'06)}, 2006, pp. 539--550.

\bibitem{aprox_evolve_1}
E.~Bergamini, H.~Meyerhenke, and C.~L. Staudt, ``Approximating betweenness
  centrality in large evolving networks,'' in \emph{Proceedings of the Meeting
  on Algorithm Engineering \& Expermiments}, ser. ALENEX '15.\hskip 1em plus
  0.5em minus 0.4em\relax Society for Industrial and Applied Mathematics, 2015,
  pp. 133--146.

\bibitem{grando_1}
F.~Grando and L.~C. Lamb, ``Estimating complex networks centrality via neural
  networks and machine learning,'' in \emph{2015 International Joint Conference
  on Neural Networks (IJCNN)}, 2015, pp. 1--8.

\bibitem{grando_2}
------, ``On approximating networks centrality measures via neural learning
  algorithms,'' in \emph{2016 International Joint Conference on Neural Networks
  (IJCNN)}, 2016, pp. 551--557.

\bibitem{survey_1}
P.~Goyal and E.~Ferrara, ``Graph embedding techniques, applications, and
  performance: A survey,'' \emph{Knowledge-Based Systems}, vol. 151, pp. 78 --
  94, 2018.

\bibitem{survey_3}
P.~Cui, X.~Wang, J.~Pei, and W.~Zhu, ``A survey on network embedding,''
  \emph{IEEE Transactions on Knowledge \& Data Engineering}, 2019.

\bibitem{harmonic}
\BIBentryALTinterwordspacing
P.~Boldi and S.~Vigna, ``Axioms for centrality,'' \emph{Internet Mathematics},
  vol.~10, no. 3-4, pp. 222--262, 2014. [Online]. Available:
  \url{https://doi.org/10.1080/15427951.2013.865686}
\BIBentrySTDinterwordspacing

\bibitem{sdne}
D.~Wang, P.~Cui, and W.~Zhu, ``Structural deep network embedding,'' in
  \emph{Proceedings of the 22Nd ACM SIGKDD International Conference on
  Knowledge Discovery and Data Mining}, ser. KDD '16.\hskip 1em plus 0.5em
  minus 0.4em\relax ACM, 2016, pp. 1225--1234.

\bibitem{gcn}
\BIBentryALTinterwordspacing
T.~N. Kipf and M.~Welling, ``Semi-supervised classification with graph
  convolutional networks,'' \emph{CoRR}, vol. abs/1609.02907, 2016. [Online].
  Available: \url{http://arxiv.org/abs/1609.02907}
\BIBentrySTDinterwordspacing

\bibitem{s2vec}
H.~Dai, B.~Dai, and L.~Song, ``Discriminative embeddings of latent variable
  models for structured data,'' in \emph{Proceedings of the 33rd International
  Conference on International Conference on Machine Learning - Volume 48}, ser.
  ICML'16.\hskip 1em plus 0.5em minus 0.4em\relax JMLR.org, 2016, pp.
  2702--2711.

\bibitem{combinatorial}
E.~Khalil, H.~Dai, Y.~Zhang, B.~Dilkina, and L.~Song, ``Learning combinatorial
  optimization algorithms over graphs,'' in \emph{Advances in Neural
  Information Processing Systems 30 (NIPS)}.\hskip 1em plus 0.5em minus
  0.4em\relax Curran Associates, Inc., 2017, pp. 6348--6358.

\bibitem{fast_bw}
M.~Riondato and E.~M. Kornaropoulos, ``Fast approximation of betweenness
  centrality through sampling,'' \emph{Data Mining and Knowledge Discovery},
  vol.~30, no.~2, pp. 438--475, Mar 2016.

\bibitem{barabasi}
A.-L. Barab{\'a}si and R.~Albert, ``Emergence of scaling in random networks,''
  \emph{Science}, vol. 286, no. 5439, pp. 509--512, 1999.

\bibitem{kendall}
\BIBentryALTinterwordspacing
M.~Kendall, \emph{Rank correlation methods}.\hskip 1em plus 0.5em minus
  0.4em\relax London: Griffin, 1948. [Online]. Available:
  \url{http://gso.gbv.de/DB=2.1/CMD?ACT=SRCHA&SRT=YOP&IKT=1016&TRM=ppn+18489199X&sourceid=fbw_bibsonomy}
\BIBentrySTDinterwordspacing

\bibitem{erdos}
P.~Erd\"{o}s and A.~R\'{e}nyi, ``On random graphs i,'' \emph{Publ. Math.
  Debrecen}, vol.~6, p. 290, 1959.

\bibitem{Watts1998Collective}
D.~J. Watts and S.~H. Strogatz, ``Collective dynamics of 'small-world'
  networks,'' \emph{Nature}, vol. 393, no. 6684, pp. 440--442, Jun. 1998.

\bibitem{snap}
J.~Leskovec and A.~Krevl, ``{SNAP Datasets}: {Stanford} large network dataset
  collection,'' \url{http://snap.stanford.edu/data}, Jun. 2014.

\bibitem{powermethod}
W.~Richards and A.~Seary, ``Eigen analysis of networks,'' \emph{Journal of
  Social Structure}, vol.~1, no.~2, pp. 1--17, 2000.

\bibitem{time_cent}
E.~C. Costa, A.~B. Vieira, K.~Wehmuth, A.~Ziviani, and A.~P.~C. Da~Silva,
  ``Time centrality in dynamic complex networks,'' \emph{Advances in Complex
  Systems}, vol.~18, no. 07n08, p. 1550023, 2015.

\bibitem{tvg_embed}
P.~Goyal, S.~R. Chhetri, and A.~Canedo, ``dyngraph2vec: Capturing network
  dynamics using dynamic graph representation learning,'' \emph{Knowledge-Based
  Systems}, 2019.

\bibitem{new_rank}
K.~Berahmand, A.~Bouyer, and N.~Samadi, ``A new local and multidimensional
  ranking measure to detect spreaders in social networks,'' \emph{Computing},
  vol. 101, no.~11, pp. 1711--1733, 2019.

\bibitem{time_evol}
C.~{Magnien} and F.~{Tarissan}, ``Time evolution of the importance of nodes in
  dynamic networks,'' in \emph{2015 IEEE/ACM International Conference on
  Advances in Social Networks Analysis and Mining (ASONAM)}, 2015, pp.
  1200--1207.

\end{thebibliography}

\end{document}